# What Colour is Neural Noise?

**J. Andrew Doyle (andrew.doyle@mcgill.ca)**
Montreal Neurological Institute, McGill University

**Alan C. Evans (alan.evans@mcgill.ca)**
Montreal Neurological Institute, McGill University

### Abstract

Random noise plays a beneficial role in cognitive processing and produces measurable improvement in simulations and biological agents' task performance. Stochastic facilitation, the phenomenon of additive noise improving signal transmission in complex systems, has been shown to occur in a variety of neural contexts. However, neuroscience analyses to date have not fully explored the colours that neural noise could be. The literature shows a $1/f$ pink noise power spectrum distribution at many levels, but many less rigourous studies assume white noise, with little justification of why that assumption is made. In this work, we briefly review the colours of noise and their useful applications in other fields. If we consider that noise is not so black and white, we could more colourfully regularize artificial neural networks and re-investigate some surprising results about how the brain benefits from noise.

**Keywords:** noise, stochastic facilitation, dithering, colour

## Introduction

In spite of the most careful and controlled data collection procedures, neural measurements are always messy. Analyzing different modalities of neural recordings always involves careful pre-processing to remove noise. Some of these sources are straightforward, such as electric power systems' induced 50/60 Hz noise present in nearly all electroencephalogram (EEG) recordings, while others are less obvious, such as magnetic susceptibility artifacts that sometimes appear in magnetic resonance imaging (MRI). Noise introduced in the measurement of neural signals should always be removed if possible for follow-up analysis.

By contrast, injecting noise is always necessary when training artificial neural networks with backpropagation. Gaussian noise has been added to training inputs (Vincent, Larochelle, Bengio, & Manzagol, 2008), injected randomly at hidden layers (Poole, Sohl-Dickstein, & Ganguli, 2014), added to weights during updates (Srivastava, Hinton, Krizhevsky, Sutskever, & Salakhutdinov, 2014), and added to the gradient to reduce sensitivity to random weight initialization (Neelakantan et al., 2015). The most successful noise-injection strategy is dropout, where during training, each artificial neuron has a Bernoulli-distributed probability of being excluded from the current training iteration (Srivastava et al., 2014). In variational autoencoders and generative adversarial networks, stochastic distributions are learned that allow networks to generate realistic data based on noise (Doersch, 2016). However, a comparison of noise with biology is difficult because in real neurons there is always a temporal component, and deep nets are not usually trained in the frequency domain to produce spike trains.

Even after removal of known artifacts, biological neural signals are always very noisy. But if noise is critical for artificial networks to learn and exhibit creativity, then how much of the biological noise is helpful? Stochastic facilitation requires noise to boost a weak periodic input signal through a non-linear system by exploiting the resonance property of some systems, where harmonic frequencies are also excited and combine to ensure signal transmission (Gammaitoni, Hänggi, Jung, & Marchesoni, 1998). Despite growing evidence that variability and noise are critical to how the brain works, most neuroscience and artificial intelligence research is stuck aggregating all noise into a single frequency, or making white/pink noise assumptions. Some of the other noise colours are shown in figure 1.

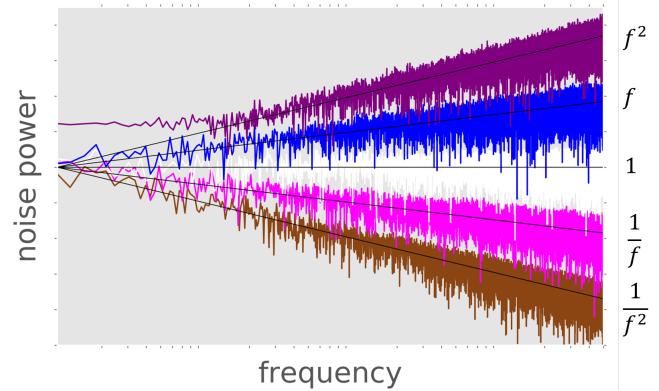

Figure 1: Power Spectral Distribution of Colours of Noise. In this log-log plot, straight lines are colours' theoretical power spectral distribution and the colours represent noise power estimated from simulated time series signals.

With analogies drawn from computer graphics and audio engineering, we hope to convince the reader to consider the possibility that the power spectral density of noise could have other colours than the pink and white noise currently in the literature.

# Noise - now in colour!

In neural decoding models, brain recordings are decoded to predict differences in individuals' performance on an experiment's task. In these models, signals and noise are often differentiated by how they are represented. In basic models, signals are deterministic linear functions of some input variables and everything else is called noise: probabilistic randomness that is unrelated to the model inputs. In more complex models, signals are also modeled as random variables that depend on the inputs but also have some stochastic variability that models our inability to learn the true, too complicated, deterministic function. These models make the distinction between fixed effects and random (noisy) effects, and the most robust analyses should model variability in the inputs as random effects as much as possible (Westfall, Nichols, & Yarkoni, 2016). Capturing all possible experimental variability in probability distributions is difficult, because investigators must make assumptions about the nature of the randomness, and it is difficult to separate the signal from the noise in the first place (McIntosh et al., 2010).

Noise colours (examples in figure 2), encode two characteristics of these assumptions: (1) the probability distribution of the amplitude, and (2) the power of noise across the frequency spectrum. Colours mostly assume Gaussian amplitude distribution and differ only in their power spectral density, $S(f)$. Different values of $\alpha$ in equation 1 determine the colour of noise, and how substantially the power of noise changes across the frequency spectrum.

$$S(f) \propto \frac{1}{f^\alpha} \qquad (1)$$

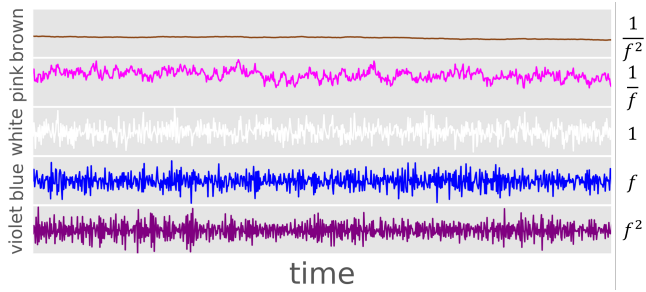

Figure 2: Time series realizations of noise colours.

Table 1 summarizes characteristics of some of the different colours of noise. The colour nomenclature for noise stems from optics, where "white" was originally described to hold equal power across the spectrum of visible light (Newton, 1718). This characterization of a flat distribution of white light has persisted even though human perception of white has been shown to diverge from colours described in physics (Thornton, 1971). Human vision has evolved to reserve more dynamic range for green than red, which is rarer in nature. To address this imbalance, digital colour spaces have been created that reserve more representation space for colour regions that we see better (Smith & Guild, 1931).

| Colour | $\alpha$ | $f$ bounds |
|---|---|---|
| Black | $> 2$ | - |
| Gray | loudness norm. | - |
| Brown | 2 | - |
| Red | 2 | *non-Gaussian |
| Pink | 1 | - |
| White | 0 | bounded |
| Green | 0 or 2 | bounded |
| Blue | $-1$ | bounded |
| Violet | $-2$ | bounded |

Table 1: Power spectrum characteristics of noise colours. Noise colours are loosely interpreted, sometimes including a range of $\alpha$. Colours with $\alpha <= 0$ must have bounds on $f$ to have finite total power.

**Black noise** has been used to describe noise with high low-frequency power ($3 < \alpha < 9$), and can be used to model rare events, including natural disasters and species extinctions (Cuddington & Yodzis, 1999).

Gray noise is described in audio engineering, and normalizes noise loudness across the spectrum of human hearing with a fourth-order polynomial function of frequency (IEC, 2013).

Brown noise power decays very rapidly with increasing frequency and is named for Brownian motion, which describes a random walk. It has been suggested that pink noise is a better model for most biological systems than brown noise (Halley, 1996).

Red noise has the same low-frequency-dominating power spectrum as brown noise, but does not always use Gaussian distributions to model amplitude (Schulz & Mudelsee, 2002). Red noise has been proposed to govern several processes in astrophysics (Do et al., 2009).

Pink noise (or "flicker" noise) is observed in many natural systems, and has been proposed as a mechanism of self-organization, as it has been argued that stable states of systems can only arise when noise can propagate infinitely (Halley, 1996). As pink noise decays with $1/f$, the lower power of high frequency noise cannot overwhelm signals and drive systems into instability (Per Bak & Wiesenfeld, 1987).

Blue noise is used for "dithering" to randomize quantization errors, which arise when representing continuous quantities discretely (example in figure 3). After dithering, black and white images can appear as if they have more intensity levels than the representation space actually supports, trading spatial resolution for intensity resolution (Hughes et al., 2014). If anyone ever prints this paper, our low resolution images will probably be dithered by the printer driver when scaling up. Although dithering can be done with any type of noise, blue noise looks best, and is the "pleasing complement of $1/f$ noise" (Ulichney, 1988).

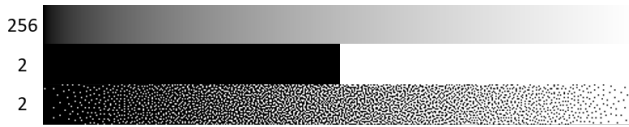

Figure 3: Gray level ramp, with number of levels in left column. (top) smooth original signal (middle) quantized signal (bottom) blue-noise dithered quantized signal.

Green noise has been proposed as an alternative dithering strategy, and consists of bounded brown (or white) noise that works particularly well on printers (Lau, Ulichney, & Arce, 2003).

Violet noise or purple noise is a good model for the ambient acoustic noise of water molecules, and has also been observed in artificial systems (Hildebrand, 2009).

Naming colours of noise is common in other fields, and despite a lack of precision on the frequency bounds and inconsistencies in exact values of $\alpha$, colours of noise are a useful tool to describe and understand a complex phenomenon in interdisciplinary communities.

## Stochastic Facilitation

Stochastic resonance is the phenomenon of Gaussian noise improving the transmission of weak signals by triggering resonance in dynamic non-linear systems (Gammaitoni et al., 1998). Brainwave entrainment is a realization of stochastic resonance, where sub-threshold periodic stimuli are presented to research subjects and shown to produce the same frequencies in the brain when noise is present. Originally described requiring white noise, the term stochastic resonance has often been more loosely used and some systems have been shown to not even require an input signal to produce periodic output (McDonnell & Abbott, 2009).

Stochastic "facilitation" has been suggested to explicitly abandon the white noise and input signal requirements and include any beneficial noise process in the brain (McDonnell & Ward, 2011). Stochastic facilitation has been observed for noisy stimuli in recordings from individual neurons (Cohen & Maunsell, 2009), electroencephelogram (EEG) (Zhou et al., 2012), magnetoencephelogram (MEG) (Linkenkaer-Hansen, Nikouline, Palva, & Ilmoniemi, 2001). In functional magnetic resonance imaging, the standard deviation of brain activation signals across trials can be used to quantify noise, and varies by region depending on the task and participants' age (Mišic, Mills, Taylor, & McIntosh, 2010). Using this measure, noise was proposed as a mechanism to handle increased uncertainty (Grady & Garrett, 2017). Despite these encouraging results, this statistic only relates a single frequency.

Noise correlation between different brain areas has also been shown to vary with attention (Cohen & Maunsell, 2009), with more information in parts of the brain with more highly correlated noise (Bejjanki, Da Silveira, Cohen, & Turk-Browne, 2017). But on the other hand, task performance improves with reduced correlation, which has striking similarities to the effect of dropout in artificial networks, which reduces the co-adaptation of neurons, and forces networks to learn many different prediction paths. This type of correlation analysis also ignores the power spectrum of the noise.

On very long time scales, the noise power spectrum distribution shifts substantially, with $\alpha$ increasing over the course of brain development (McIntosh et al., 2010). But very short time scales may have been overlooked: EEG/MEG pre-processing involves low-pass filtering, which removes high frequencies. This might make detecting weak signals in the data possible at the cost of removing beneficial noise or its harmonics.

Stochastic facilitation has been shown in signal processing and physics to perform dithering (Wannamaker, Lipshitz, & Vanderkooy, 2000), which looks even better with blue noise. Mammalian-inspired visual pathways can be designed to produce dithering (Masmoudi, Antonini, & Kornprobst, 2013), and perturbing individual neurons' spike trains constitutes dithering (Pazienti, Maldonado, Diesmann, & Grün, 2008), but none have yet detected blue noise dithering in neuroscience.

## Conclusion

Despite clues from other fields, we do not yet have any convincing evidence that noise in the brain is more colourful than what has been described to date in the literature. But the concept that noise might have a variable power spectral density could be employed to help prediction generalization in artificial neural networks by adding coloured noise to inputs as a form of data augmentation.

In this work, we have introduced a way to more easily discuss different power spectral densities in terms of colour instead of mathematics. By reviewing the successful models and beneficial properties of noise outside of neuroscience, we have raised the possibility that noise might not be all pink and white in the brain, and that blue-noise dithering might play an important role in cognition, filling in missing information stochastically at higher frequencies.